# Tailoring Defects in Photonic Time Crystals for Coherent Energy Control

Dayeong Lee, Jongheon Yeo, Gitae Lee, Jungmin Kim, Namkyoo Park, and Sunkyu Yu

***Abstract*—Recent advances in time-varying photonics have revealed new degrees of freedom for manipulating optical states, arising from the distinctive nature of the temporal axis: causality and open-system dynamics. A representative example is photonic time crystals (PTCs) characterized by discrete time-translational symmetry, which exhibit space-analogous yet distinct phenomena, such as momentum gaps and amplifying-decaying Floquet-mode pairs. Although PTCs enable optical-energy amplification beyond conventional gain media, their application as programmable energy-functional devices remains challenging. Here, we propose a design framework for tailoring optical energy via defective PTCs. By optimizing defect permittivity and duration using analytic gradients of time transfer matrices, we realize prescribed coherent energy amplification and suppression. We show that a single defect enables continuous energy tailoring, while revealing an intrinsic asymmetry between amplification and suppression due to the inherently amplifying nature of the momentum gap. Extending the framework to coupled defects expands the design space and markedly improves suppression, establishing temporal-defect engineering as a route to programmable coherent energy control.**

## I. INTRODUCTION

DRIVEN by the need for additional degrees of freedom in manipulating optical states, time-varying photonics has been intensively studied not only for exploring novel material phases defined along the temporal axis [1-11], but also for addressing scalability challenges in optical computations [12-14]. One representative example is the study of light behavior in photonic time crystals (PTCs) [2, 15, 16], which are defined by discrete time-translational symmetry and support momentum gaps as temporal counterparts of the energy gaps in spatial photonic crystals [17]. In contrast to spatial photonic crystals, where time-translation symmetry conserves frequency under lossless unitary evolution, PTCs break this symmetry and enable energy exchange with the external modulation drive. This produces momentum gaps in which Floquet modes emerge as causality-constrained amplifying and decaying pairs [2, 16]. Consequently, one representative signature of PTCs is the amplification of optical energy within their momentum gap.

To impose richer control over optical energy beyond monotonic amplification, it is necessary to break discrete translational symmetry, analogous to guiding and resonances within bandgaps of spatial photonic crystals [17]. Such symmetry breaking enables temporal defects to reshape the interaction between a pair of Floquet modes inside the momentum gap, thereby providing a route toward programmable energy enhancement or suppression. However, previous studies have mainly explored statistically homogeneous breaking of discrete translational symmetry through uncorrelated [3] or correlated [4] temporal disorder, or accidental emergence of scattering and localization effects in defective PTCs [18, 19]. A systematic design framework for temporal defects that realize prescribed optical-energy responses has remained challenging.

Here, we develop an energy-objective design framework for temporal defects in finite PTCs by optimizing defect permittivity and duration through analytic gradients of the time-ordered transfer matrix. Using a single defect, we reveal an intrinsic asymmetry in energy tailoring: coherent suppression is more challenging due to the inherently amplifying nature of the momentum gap. We extend our design framework to a double-defect system, where the forward-backward wave ratio acts as an effective boundary condition for reciprocal-like defect interactions despite temporal causality. These interactions broaden the design space, degrading optimization stability for amplification but markedly improving suppression, demonstrating functionality-dependent defect-number design. Our results provide a general recipe for optical devices with tailored energy responses based on temporal modulation, including lasers, amplifiers, and perfect absorbers.

## II. DESIGN FRAMEWORK

### A. *Temporal transfer-matrix formulation*

We consider a spatially homogeneous and temporally finite PTC with period $T_0$, as illustrated in Fig. 1(a). The unit cell of the PTC consists of two distinct temporal layers with relative permittivities $\varepsilon_1$ and $\varepsilon_2$. We focus on an operating bandwidth in which the constituent materials are assumed to be linear, lossless, and dispersionless. For $x$-polarized light with the

This work was supported by the National Research Foundation of Korea (NRF) through the Innovation Research Center (No. RS-2024-00413957), Core Research Grants (No. RS-2026-25469085), Pilot and Feasibility Grants (No. RS-2025-19912971), Midcareer Researcher Program (No. RS-2023-00274348), and Young Researcher Infrastructure Program (No. RS-2024-00398604), all funded by the Korean government. This work was supported by Creative-Pioneering Researchers Program and the BK21 FOUR program of the Education and Research Program for Future ICT Pioneers in 2026, through Seoul National University. We also acknowledge administrative support from SOFT foundry institute. *(Dayeong Lee and Jongheon Yeo contributed equally to this work.) (Corresponding authors: Jungmin Kim, Namkyoo Park and Sunkyu Yu.)*

Dayeong Lee, Jongheon Yeo, Gitae Lee, Jungmin Kim, Namkyoo Park, and Sunkyu Yu are with the Department of Electrical and Computer Engineering, Seoul National University, Seoul 08826, Republic of Korea (e-mail: dayeong315@snu.ac.kr; zkdnqh12@snu.ac.kr; ltk175@snu.ac.kr; jmkim93@gmail.com; nkpark@snu.ac.kr; sunkyu.yu@snu.ac.kr).

displacement field $\mathbf{D}(\mathbf{r}, t) = \mathbf{e}_x D(t)e^{ikz}$, the governing equation is given by $[d^2/dt^2 + c^2k^2/\varepsilon(t)]D(t) = 0$ [4, 20], where $c$ is the speed of light, $k$ is the wavenumber, and $\varepsilon(t)$ is the time-varying relative permittivity.

In the PTC, we introduce temporal defects—defined as temporal intervals whose permittivity and duration differ from those of the PTC layers—embedded between PTC unit cells, as shown in Fig. 1(b). Each defect is characterized by a pair of parameters, $\mathbf{d}_l = (\varepsilon_{d,l}, t_{d,l}/T_0)$, where $\varepsilon_{d,l}$ and $t_{d,l}$ denote the relative permittivity and duration of the $l$-th defect, respectively.

The optical state in the PTC is represented as $|\Psi(t)\rangle = [\Psi_f(t), \Psi_b(t)]^T$, where $\Psi_f$ and $\Psi_b$ denote the forward ($+z$) and backward ($-z$) propagating components of the displacement field, as $D(t) = \Psi_f(t) + \Psi_b(t)$. To analyze the evolution of $|\Psi(t)\rangle$ through the defective PTC, we employ the temporal transfer-matrix formulation [21, 22]. A PTC containing $L$ defects ($l = 1, 2, \ldots, L$) can possess $L + 1$ pristine PTC segments. The transfer matrices through the $s$-th PTC segment containing $N_s$ unit cells ($s = 1, 2, \ldots, L + 1$) and through the $l$-th defect are defined as $M_s$ and $F_l = F(\mathbf{d}_l)$, respectively (Fig. 1(c)). For the input ($|\Psi_{in}\rangle$) and output states ($|\Psi_{out}\rangle$) (Fig. 1(b)), the total transfer matrix for $|\Psi_{out}\rangle = M(\{\mathbf{d}_l\})|\Psi_{in}\rangle$ is written as

$$M(\{\mathbf{d}_l\}) = A_F M_{L+1} F_L M_L \cdots F_2 M_2 F_1 M_1, \quad (1)$$

where each $M_s$ connects the state immediately before the $s$-th PTC segment to the state at the end of the segment, and $A_F$ denotes the final interface matrix (Supplementary Note S1 for details). While the corresponding pristine PTC contains a total of $N_{total} = \sum_{s=1}^{L+1} N_s$ unit cells, we impose $N_I \triangleq N_1 \geq 1$ and $N_F \triangleq N_{L+1} \geq 1$, with $N_s \geq 0$ for $s = 2, \ldots, L$. This ensures that all defects lie inside the PTC while allowing adjacent defects with $N_s = 0$.

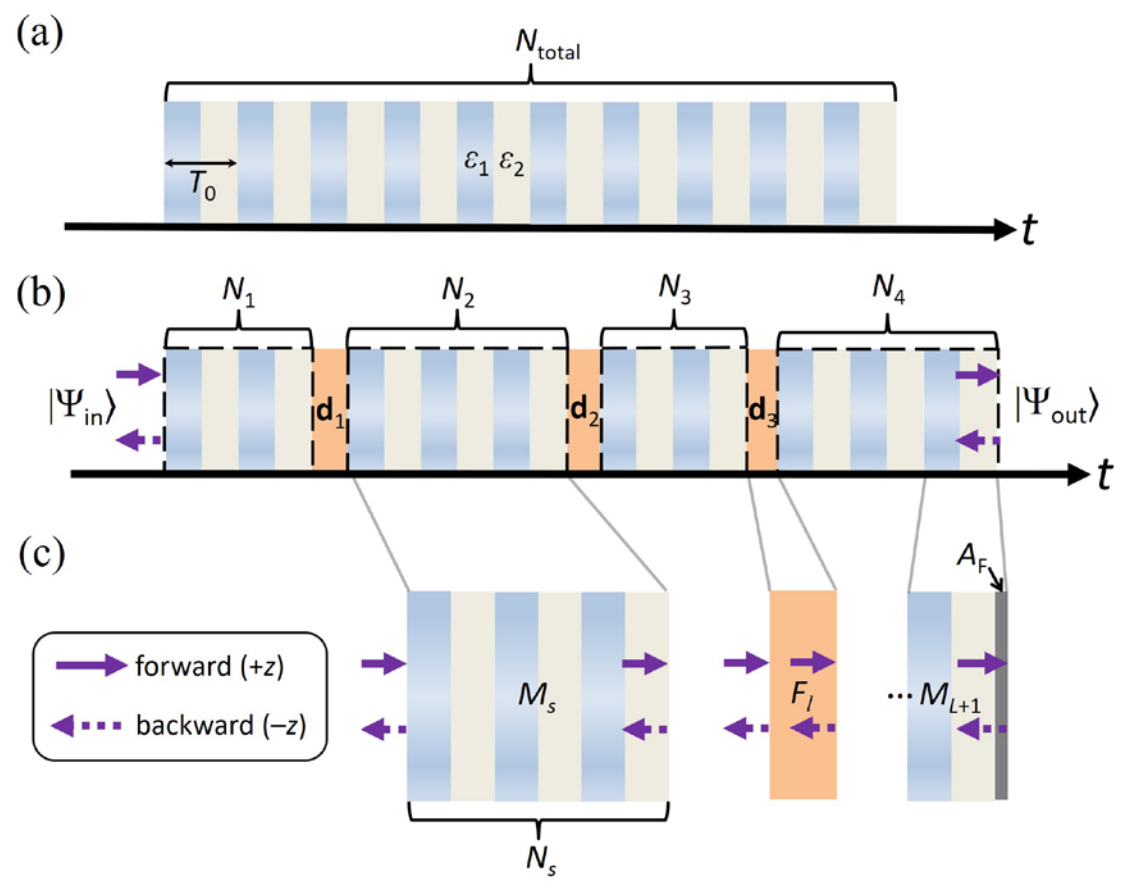

**Fig. 1.** Defective PTC. (a,b) Schematics of a finite pristine PTC (a) and defective PTC (b). (c) Transfer matrices for building blocks. In (b,c), solid and dashed arrows denote forward ($+z$) and backward ($-z$) propagating components, respectively. The pristine PTC has $\varepsilon_1 = 3$ and $\varepsilon_2 = 1$ throughout this work.

### *B. Energy objective and optimization protocol*

To design a defective PTC for tailoring energy of interacting light, we define the cost function as

$$C(\{\mathbf{d}_l\}) = \frac{1}{2}\left(\frac{E_{out}(\{\mathbf{d}_l\})}{E_{in}} - \rho\right)^2, \quad (2)$$

where $E_{in} = \langle\Psi_{in}|\Psi_{in}\rangle$ and $E_{out} = \langle\Psi_{out}|\Psi_{out}\rangle$ are the electromagnetic energies of the input and output states, respectively, and $\rho \geq 0$ is the target output-to-input energy ratio: $\rho < 1$ for energy suppression and $\rho > 1$ for energy amplification.

Let $d_{l,n}$ ($n = 1, 2$) denote the $n$-th component of the defect parameter vector $\mathbf{d}_l$ with $d_{l,1} = \varepsilon_{d,l}$ and $d_{l,2} = t_{d,l}/T_0$. The gradient of $C$ with respect to the alteration of defects becomes

$$\frac{\partial C}{\partial d_{l,n}} = \frac{2}{E_{in}}\left(\frac{E_{out}}{E_{in}} - \rho\right)\mathrm{Re}\left[\left\langle \Psi_{out}\left|\frac{\partial M}{\partial d_{l,n}}\right|\Psi_{in}\right\rangle\right]. \quad (3)$$

In (3), the partial derivative of $M$ is obtained as

$$\frac{\partial M}{\partial d_{l,n}} = \frac{\partial}{\partial d_{l,n}}\left(M_{>l} F_l M_{<l}\right) = \frac{\partial M_{>l}}{\partial d_{l,n}} F_l M_{<l} + M_{>l}\frac{\partial F_l}{\partial d_{l,n}} M_{<l}, \quad (4)$$

where $M_{<l}$ and $M_{>l}$ denote the matrix products before and after the $l$-th defect in time, respectively. The $\partial M_{>l}/\partial d_{l,n}$ term—the dependence of the following temporal interface on the defect permittivity—can be regrouped into a local defect-layer matrix (Supplementary Note S2 for details). The computational cost for calculating (3) is $O(L)$ by precomputing each $M_s$.

Under the constrained parameter space $D = \{\varepsilon_{min} \leq \varepsilon_{d,l} \leq \varepsilon_{max}, \tau_{min} \leq t_{d,l}/T_0 \leq \tau_{max}\}$, we optimize the defect parameters by the following gradient descent approach, $d_{l,n}^{(j+1)} = \prod_D[d_{l,n}^{(j)} - \eta_n \partial C/\partial d_{l,n}]$, where $j$ is the iteration index, $\eta_1 = \eta_\varepsilon$ and $\eta_2 = \eta_\tau$ are the learning rates for the first and second components of $\mathbf{d}_l$, respectively, and $\prod_D$ denotes the clipping onto $D$. For each case, we test 100 random initializations under the same parameter bounds.

## III. Results

### *A. Single-defect design*

In manipulating optical energy with defective PTCs, we focus on coherent input with a deterministic phase difference between the forward and backward waves in the input state, using $|\Psi_{in}\rangle = [1, 1]^T/2^{1/2}$ throughout this work. The wavenumber $k$ inside the entire system is set to lie within the momentum gap of the pristine PTC, in order to exploit imaginary parts of eigenfrequencies for energy manipulation, such as coherent suppression and amplification [23].

We begin with the simplest class of defective PTCs: a single-defect configuration with $L = 1$, $\varepsilon_d \triangleq \varepsilon_{d,1}$, and $t_d \triangleq t_{d,1}$. For a fixed number of pristine PTC unit cells, $N_{total} = 10$, the position of the defect is specified by $N_I$. Accordingly, for a given $N_I$, the defect state is defined in the two-dimensional (2D) design space ($\varepsilon_d$, $t_d/T_0$). Figure 2(a) shows the cost-function landscape in ($\varepsilon_d$, $t_d/T_0$) for the complete-suppression target ($\rho = 0$) at $N_I = 5$. Although the landscape is nonconvex with multiple local minima, our gradient-descent approach successfully converges to one of several local minima, depending on the learning parameters (Fig. 2(b)). This successful convergence persists over a broad range of $\rho$ (Fig. 2(c)), covering both coherent suppression ($\rho << 1$) and amplification ($\rho >> 1$) regimes. Therefore, as shown in Fig. 2(d) for the normalized energy evolution $E(t)/E_{in}$, the defective

PTC enables switching between substantial suppression (blue line) and amplification (orange line) by altering only a single defect while leaving the pristine unit cells unchanged.

As another critical factor affecting optimization performance, we examine the effect of the defect position within the PTC, as shown in Fig. 2(e). Although the coherent-amplification target is stably achieved by our design approach (orange line), coherent suppression depends strongly on the defect position, characterized by $N_{\mathrm{I}}$, with the suppression performance degrading rapidly as $N_{\mathrm{I}}$ increases. This result indicates that, within the momentum gap of the pristine PTC, where the eigenfrequencies acquire imaginary parts with opposite signs, amplification intrinsically dominates the wave evolution, as illustrated on the left side of the defect in Fig. 2(d). Therefore, for the suppression target, the defect enhances the decaying modes through designed interference; this requires sufficient remaining propagation time to fully suppress the optical energy amplified before the wave reaches the defect.

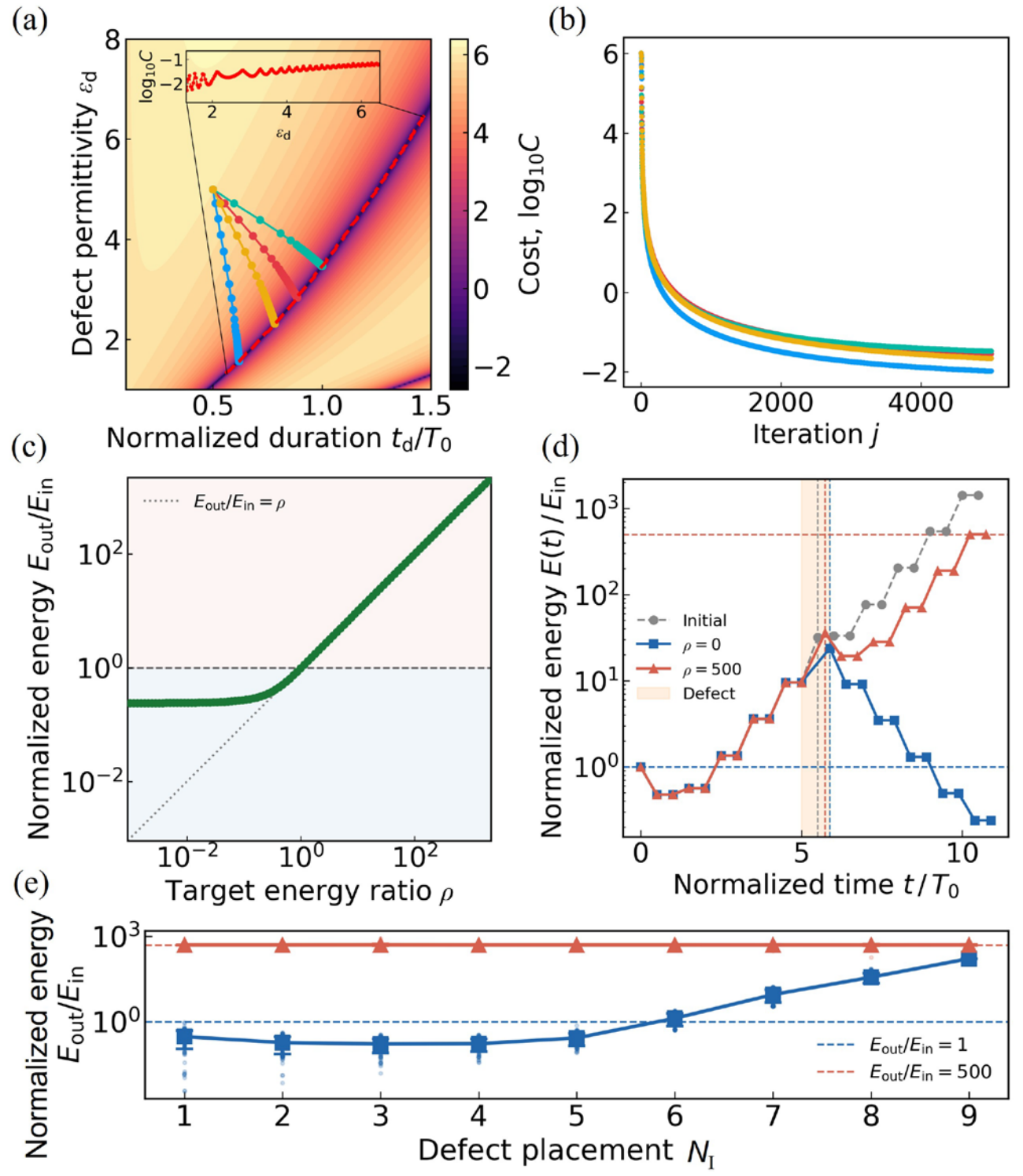

**Fig. 2.** Inverse design of a single defect. (a) Cost landscape with trajectories. $\eta_\varepsilon = 1 \times 10^{-6}$, $\eta_\tau = \{0.5, 1.5, 2.5, 4.5\} \times 10^{-8}$ (blue, yellow, red, and green solid lines). Red dashed line denotes cost minima. (b) Cost evolution of (a). (c) Optimized $E_{\mathrm{out}}/E_{\mathrm{in}}$ versus $\rho$. (d) $E(t)/E_{\mathrm{in}}$ for the initial, $\rho = 0$, and 500 cases. Shaded regions with vertical lines indicate defects. (e) Optimized $E_{\mathrm{out}}/E_{\mathrm{in}}$ for $\rho = 0$ (blue) and 500 (orange). Squares and triangles denote averages, and error bars denote standard deviations over 100 initializations. (a-d) $N_{\mathrm{I}} = 5$. (d,e) Horizontal dashed lines denote the values of 1 (blue) and 500 (orange). (c-e) $\eta_\varepsilon = 1 \times 10^{-6}$ and $\eta_\tau = 2.5 \times 10^{-8}$. $\varepsilon_{\mathrm{min}} = 1$, $\varepsilon_{\mathrm{max}} = 8$, $\tau_{\mathrm{min}} = 0.1$, and $\tau_{\mathrm{max}} = 1.5$. $k = 0.6\omega_0/c$ with $\omega_0 = 2\pi/T_0$.

### B. *Design of interacting defects*

Beyond the single-defect design, we extend our method to multi-defect configurations. Specifically, we consider a PTC with two defects, characterized by four continuous variables, ($\varepsilon_{\mathrm{d},1}$, $t_{\mathrm{d},1}/T_0$, $\varepsilon_{\mathrm{d},2}$, $t_{\mathrm{d},2}/T_0$). With the number of pristine PTC unit cells fixed at $N_{\mathrm{total}} = 10$, the defect positions are specified by $N_{\mathrm{I}}$ and $N_{\mathrm{F}}$, satisfying $N_{\mathrm{I}} + N_{\mathrm{F}} \leq 10$. Because we assume the hard-core constraint that prohibits temporal overlap between defects, $N_{\mathrm{I}} + N_{\mathrm{F}} = 10$ corresponds to a configuration in which a single contiguous defect is composed of two adjacent segments. Figure 3(a) shows representative examples of the energy evolution for coherent suppression (blue line) and amplification (orange line), demonstrating that the monotonic evolution between the defects is altered by their interaction.

In Fig. 3(b), we show the statistics of the suppression and amplification performance obtained from 100 random initializations of the defect parameters for every ($N_{\mathrm{I}}$, $N_{\mathrm{F}}$), including a comparison between the single- and double-defect configurations. For the amplification target, which is well aligned with the inherent wave dynamics in the momentum gaps, the design objective is successfully achieved in both cases, with $0.90\rho \leq E_{\mathrm{out}}/E_{\mathrm{in}} \leq 1.10\rho$ attained for 99.4% and 98.3% of the initializations in the single- and double-defect configurations, respectively. However, the double-defect configuration exhibits degraded design stability, which may be attributed to an increased number of local minima arising from the larger number of design parameters. This degradation in optimization stability is also reflected in the position-dependent performance (Fig. 3(c)).

In contrast, the expanded design space improves the performance for coherent suppression, which is a more challenging task in defective PTCs. Figure 3(b) shows that the double-defect configuration enhances the suppression performance, increasing the fraction of cases with $E_{\mathrm{out}}/E_{\mathrm{in}} < 1$ from 59.3% to 68.3%. This improvement becomes more pronounced for deeper suppression, with the fraction satisfying $E_{\mathrm{out}}/E_{\mathrm{in}} < 10^{-1}$ increasing from 10.3% to 38.0% (Supplementary Note S3 for details). Notably, although the position dependence of the suppression performance becomes more intricate as a function of ($N_{\mathrm{I}}$, $N_{\mathrm{F}}$), the overall trend is maintained: the suppression performance is substantially degraded for large $N_{\mathrm{I}}$ (Fig. 3(d)), owing to the inherent amplification of light in the pristine PTC.

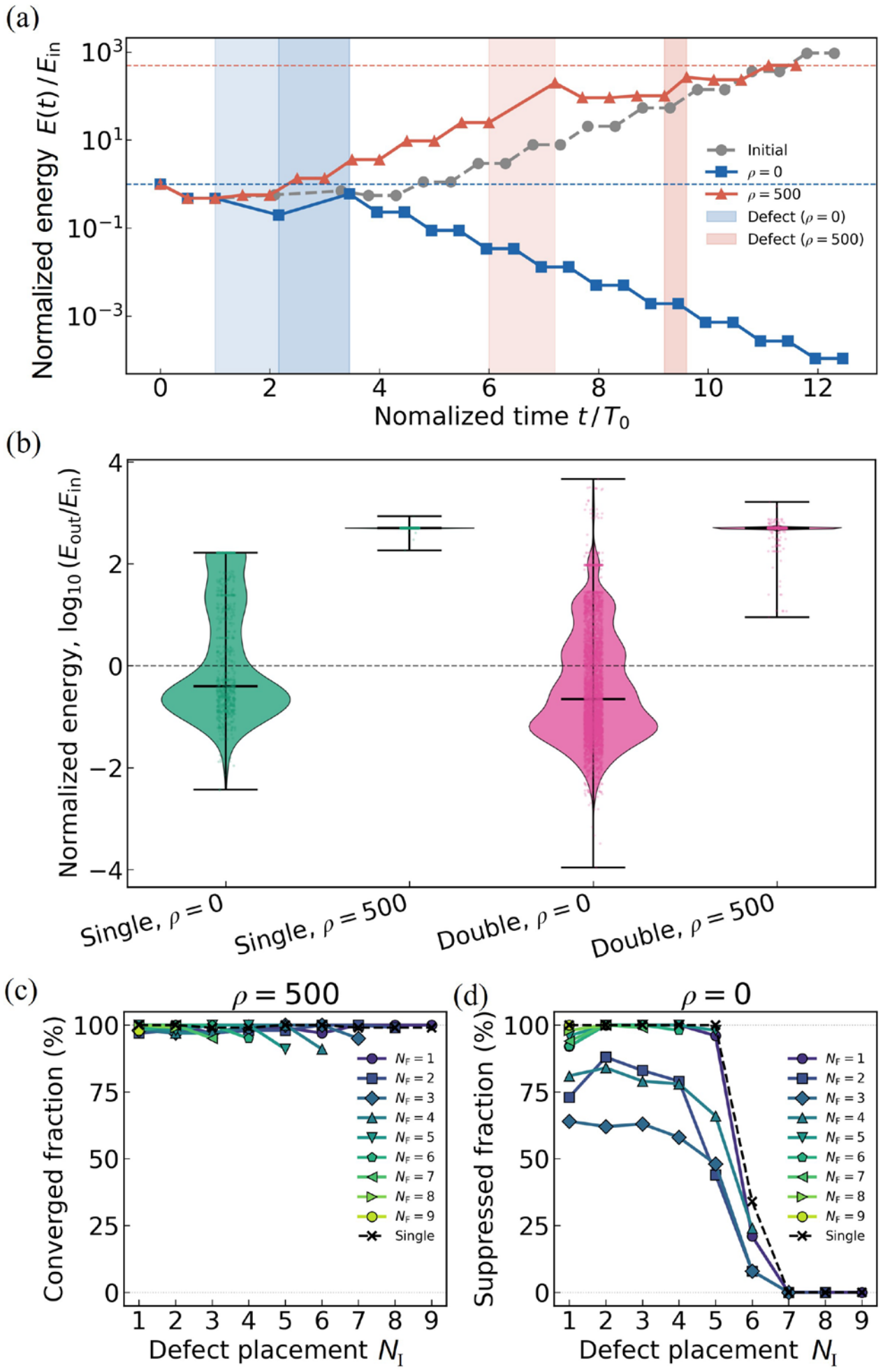

**Fig. 3.** Double-defect design. (a) Representative energy evolutions for the initial, suppression-optimized ($\rho$ = 0), and amplification-optimized ($\rho$ = 500) cases. ($N_I$, $N_F$) = (1, 9) for $\rho$ = 0 and ($N_I$, $N_F$) = (6, 2) for $\rho$ = 500. Shaded regions indicate defects. Dashed lines denote the normalized energy of 1 (blue) and 500 (orange). (b) Normalized output energy distributions of single- and double-defect configurations, with width indicating probability density. (c,d) Converged fraction (c) and suppressed fraction (d) with varying ($N_I$, $N_F$) and dashed lines for single-defect case. $\eta_\varepsilon = 1 \times 10^{-5}$ and $\eta_\tau = 5 \times 10^{-8}$. 100 random initializations for each ($N_I$, $N_F$). All other parameters are the same as those in Fig. 2.

### C. *Effect of defect interactions*

Motivated by the hybridization of spatial defect modes [17], which depends on the spatial separation between the modes, we examine the underlying physics in terms of the inter-defect separation $N_M \triangleq N_2$ for both the suppression and amplification targets. We focus on the distributions of the optimized defect permittivities and durations, which are essential for practical implementation. The distributions are evaluated over valid solutions defined according to each target: $E_{out}/E_{in} < 1$ for $\rho = 0$ and $0.90\rho \leq E_{out}/E_{in} \leq 1.10\rho$ for $\rho = 500$ (Supplementary Note S3 for $E_{out}/E_{in} < 10^{-1}$, $10^{-2}$, and $10^{-3}$ cases).

Figures 4(a) and 4(b) show the distributions of the single-defect solution parameters in the ($\varepsilon_d$, $t_d/T_0$) plane. For comparison, we evaluate the corresponding distributions for double-defect systems with large ($N_M$ = 5: Figs. 4(c,d)) and zero ($N_M$ = 0: Figs. 4(e,f)) inter-defect separations. Notably, while the single-defect case exhibits narrow, branch-like distributions, the double-defect cases show much broader distributions. This broadening becomes more pronounced as $N_M$ decreases, whereas the larger-separation case ($N_M$ = 5) retains remnants of the single-defect branches.

This trend is quantified by the grid-occupancy coverage in Fig. 4(g), defined as the fraction of discretized parameter-space cells containing at least one valid solution. This metric represents the occupation of the solution space in the allowed parameter domain. For both the suppression and amplification targets, the coverage decreases as $N_M$ increases, approaching the single-defect reference values at sufficiently large separations, analogous to the decoupling of hybridized modes in spatial defect systems. Therefore, the broad solution set at small $N_M$ reflects strong inter-defect interactions, which further expand the design space for achieving the target performance. From the perspective of one-to-many relationships between optical responses and material configurations [24], the effective interactions between temporal defects enlarge the set of allowed temporal-defect configurations that realize the target optical functionality.

Although the temporal defect interactions are analogous to hybridization in spatial defects, their underlying mechanisms are distinct due to causality. In contrast to spatial defects, which can couple reciprocally through the overlap of localized modes, temporal defects interact only through forward causal evolution. Nevertheless, because the relative amplitudes and phases of the forward- and backward-wave components define an effective boundary condition, the coupled response of temporal defects can be modeled within the same reciprocal governing framework. The inter-defect separation $N_M$ determines the strength of this effective interaction and therefore controls the collective response of the temporal defects.

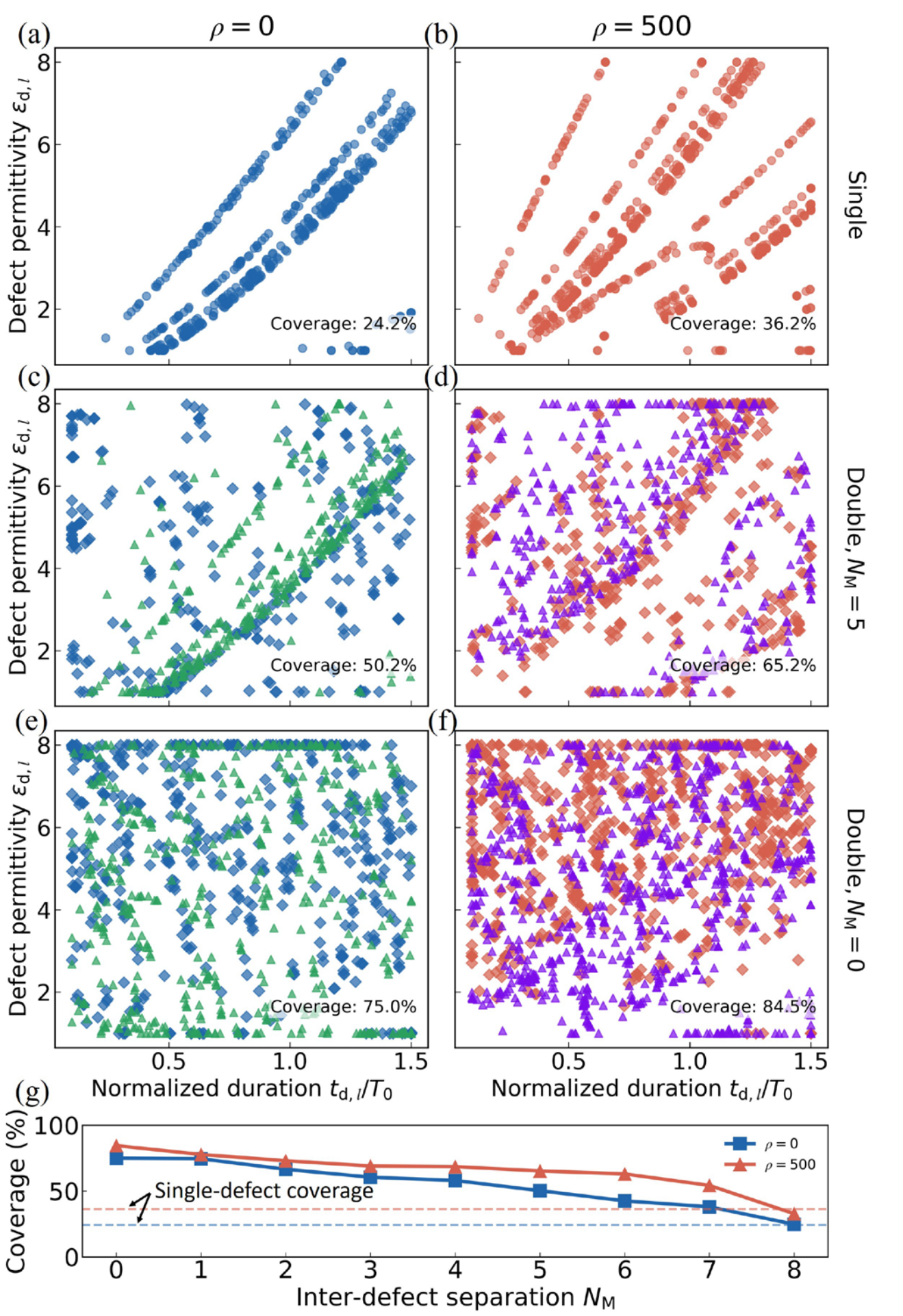

**Fig. 4.** Solution space under $\rho = 0$ and 500. (a,b) Single-defect case. (c,d) Double-defect case at $N_M = 5$. (e,f) Double-defect case at $N_M = 0$. (c-f) $\mathbf{d}_1$ (triangles) and $\mathbf{d}_2$ (diamonds) are shown separately. Grid-occupancy coverages are indicated in each panel. (g) Grid-occupancy coverage (%) for $\rho = 0$ (blue squares) and $\rho = 500$ (orange triangles). Dashed lines indicate the single-defect coverage. All other parameters are the same as those in Fig. 3. 20 × 20 grid discretization for coverage evaluation.

## IV. Discussion

The present framework demonstrates programmable control of optical energy, but several extensions remain. In our design, we assumed linear, lossless, and dispersionless media with instantaneous permittivity switching. Incorporating dispersion, finite modulation speed, and loss or gain through generalized transfer matrices would allow for handling more realistic platforms, although the enlarged parameter space may constrain experimentally accessible designs. Scaling the framework to many defects would also increase the non-convexity of the optimization landscape, motivating global or hybrid optimization strategies beyond gradient descent [25, 26].

Another promising physical direction is to treat the input coherence state as a design variable [23]. In the present work, we fixed the amplitudes and phases of the forward and backward waves, which sets the effective boundary condition for the temporal-defect response. Joint optimization of the input coherence state and defect parameters could expand the set of temporal configurations capable of realizing a prescribed energy response. Extending the framework to incoherent illumination would require replacing the state-based objective with an ensemble-averaged or correlation-matrix formulation, in line with the growing interest in incoherent light in photonics [27-29].

In conclusion, we established an inverse-design framework for temporal defects in finite PTCs. By formulating coherent suppression and amplification within a unified energy-based objective function and demonstrating that coupled temporal defects expand the accessible solution space, this work provides a route toward programmable energy control in time-varying photonic media.

# Supplementary Information for "Tailoring Defects in Photonic Time Crystals for Coherent Energy Control"

Dayeong Lee[1*], Jongheon Yeo[1*], Gitae Lee[1], Jungmin Kim[1,2‡], Namkyoo Park[2§], and Sunkyu Yu[1†]

[1]Intelligent Wave Systems Laboratory, Department of Electrical and Computer Engineering, Seoul National University, Seoul 08826, Korea

[2]Photonic Systems Laboratory, Department of Electrical and Computer Engineering, Seoul National University, Seoul 08826, Korea

E-mail address for correspondence: ‡jmkim93@gmail.com, §nkpark@snu.ac.kr, †sunkyu.yu@snu.ac.kr

*These authors contributed equally to this work.

**Supplementary Note S1. Transfer-matrix formulation**

We use a transfer matrix formalism to relate the input and output states of a defective PTC, as $|\Psi_{\text{out}}\rangle = M|\Psi_{\text{in}}\rangle$:

$$M = A_{\text{ext}}{}^{-1} T_N T_{N-1} \cdots T_1 A_{\text{ext}}, \quad T_m = A_m B_m A_m^{-1} \quad (m = 1, 2, \cdots, N), \tag{S1}$$

where $N$ is the total number of temporal layers in the finite medium, $m$ denotes layer index, $A_m$ and $B_m$ denote the interface and propagating matrices of the $m$-th layer, and $A_{\text{ext}}$ is the conversion matrix for the external medium. The transfer matrix of the $m$-th layer is

$$\begin{aligned} T_m = A_m B_m A_m{}^{-1} &= \begin{pmatrix} 1 & 1 \\ Z_m & -Z_m \end{pmatrix} \begin{pmatrix} e^{-i\omega_m \Delta_m} & 0 \\ 0 & e^{i\omega_m \Delta_m} \end{pmatrix} \begin{pmatrix} 1 & 1 \\ Z_m & -Z_m \end{pmatrix}^{-1} \\ &= \begin{pmatrix} \cos(\omega_m \Delta_m) & -iZ_m{}^{-1} \sin(\omega_m \Delta_m) \\ -iZ_m \sin(\omega_m \Delta_m) & \cos(\omega_m \Delta_m) \end{pmatrix}, \end{aligned} \tag{S2}$$

where $k$ is the wavenumber, $Z_m = (\mu_0/\varepsilon_0)^{1/2}\varepsilon_m{}^{-1/2}$ is the impedance, $\varepsilon_m$ is the relative permittivity, $\omega_m = kc/\varepsilon_m{}^{1/2}$ is the angular frequency, and $\Delta_m$ is the duration of the $m$-th layer.

The matrices $M_s$ in the main text for $s = 1, 2, \ldots, L + 1$ connect the optical state immediately before the $s$-th PTC segment to the state at the end of the segment. For example, a single-defect system with two PTC unit cells, containing two layers for each unit cell, can be written as $M = A_F M_2 F_1 M_1$, with $s = 1, 2$ and $l = 1$. Then, $M_1 = B_2 A_2{}^{-1} A_1 B_1 A_1{}^{-1} A_{\text{ext}}$, $M_2 = B_5 A_5{}^{-1} A_4 B_4 A_4{}^{-1} A_3(\varepsilon_d)$, and the defect matrix $F_1 = B_3(k, \varepsilon_d, t_d/T_0) A_3{}^{-1}(\varepsilon_d) A_2$ is inserted between PTC segments. The interface matrix $A_F$ in the main text becomes $A_F = A_{\text{ext}}{}^{-1} A_N$. For multiple defects, the same grouping leads to Eq. (1) of the main text.

In contrast to the layer index $N$, the quantities $N_s$ and $N_{\text{total}}$ used in the main text count the numbers of PTC unit cells. For the PTC with binary material phases considered in the main text, each modulation period consists of two layers, and each defect corresponds to an additional layer, leading to $N = 2N_{\text{total}} + L$.

**Supplementary Note S2. Derivation of the gradient**

We derive the gradient of the transfer matrix. For intuition, we start from the example of a single-defect system in Note S1, $F_1 = B_3(k, \varepsilon_d, t_d/T_0)A_3^{-1}(\varepsilon_d)A_2$, $M_{>1} = A_F M_2 = A_F B_5 A_5^{-1} A_4 B_4 A_4^{-1} A_3(\varepsilon_d)$, and $M_{<1} = M_1 = B_2 A_2^{-1} A_1 B_1 A_1^{-1} A_{ext}$, noting that $M_{>1}$ also depends on $\mathbf{d}_1$ whereas $M_{<1}$ does not. The product rule applied to $M = M_{>1} F_1 M_{<1}$ yields $(\partial M_{>1}/\partial d_n)F_1 M_{<1} + M_{>1}(\partial F_1/\partial d_n)M_{<1}$, where $d_n \in \{\varepsilon_d, t_d/T_0\}$ denotes a component of $\mathbf{d}_1$, giving the same form of Eq. (4) in the main text.

However, for evaluating the gradients, it is convenient to regroup all the parameters of the defect of interest into a single transfer matrix $T_d \triangleq T_m = A_d B_d A_d^{-1}$, where $m$ is the layer index of the defect with $A_d \triangleq A_m$ and $B_d \triangleq B_m$. Following Eq. (S1), $M$ can be factorized to be

$$M = A_{ext}^{-1} P_F T_d P_I A_{ext}, \tag{S3}$$

where $P_I \triangleq T_{m-1} \cdots T_1$ and $P_F \triangleq T_N \cdots T_{m+1}$ denote the products before and after $T_d$ in time, respectively. $P_I$ and $P_F$ contain only pristine PTC layers in the single-defect case, and additionally include all other defect layers except for the layer $T_d$ for multi-defect systems. The local matrix derivatives for defect parameters of interest are

$$\begin{aligned}
\frac{\partial A_d}{\partial \varepsilon_d} &= \begin{pmatrix} 0 & 0 \\ -\dfrac{Z_d}{2\varepsilon_d} & \dfrac{Z_d}{2\varepsilon_d} \end{pmatrix}, \\
\frac{\partial B_d}{\partial \varepsilon_d} &= -\frac{\omega_d}{2\varepsilon_d} \begin{pmatrix} -it_d e^{-i\omega_d t_d} & 0 \\ 0 & it_d e^{i\omega_d t_d} \end{pmatrix}, \\
\frac{\partial B_d}{\partial (t_d / T_0)} &= T_0 \begin{pmatrix} -i\omega_d e^{-i\omega_d t_d} & 0 \\ 0 & i\omega_d e^{i\omega_d t_d} \end{pmatrix},
\end{aligned} \tag{S4}$$

where $Z_d = (\mu_0/\varepsilon_0)^{1/2}\varepsilon_d^{-1/2}$ is the impedance and $\omega_d = kc/\varepsilon_d^{1/2}$ is the angular frequency of $T_d$. Using the product rule and $\partial(A^{-1}) = -A^{-1}(\partial A)A^{-1}$, we obtain

$$\begin{aligned}
&\frac{\partial T_\mathrm{d}}{\partial(t_\mathrm{d}/T_0)} = A_\mathrm{d}\frac{\partial B_\mathrm{d}}{\partial(t_\mathrm{d}/T_0)}A_\mathrm{d}^{-1},\\
&\frac{\partial T_\mathrm{d}}{\partial\varepsilon_\mathrm{d}} = \frac{\partial A_\mathrm{d}}{\partial\varepsilon_\mathrm{d}}B_\mathrm{d}A_\mathrm{d}^{-1} + A_\mathrm{d}\frac{\partial B_\mathrm{d}}{\partial\varepsilon_\mathrm{d}}A_\mathrm{d}^{-1} - A_\mathrm{d}B_\mathrm{d}A_\mathrm{d}^{-1}\frac{\partial A_\mathrm{d}}{\partial\varepsilon_\mathrm{d}}A_\mathrm{d}^{-1}.
\end{aligned} \tag{S5}$$

Combining Eq. (S3) and Eq. (S5) gives $\partial M/\partial d_n = A_\mathrm{ext}^{-1}P_\mathrm{F}(\partial T_\mathrm{d}/\partial d_n)P_\mathrm{I}A_\mathrm{ext}$. Because the input state is fixed, the partial derivative of $|\Psi_\mathrm{out}\rangle = M|\Psi_\mathrm{in}\rangle$ is obtained as $\partial|\Psi_\mathrm{out}\rangle/\partial d_n = (\partial M/\partial d_n)|\Psi_\mathrm{in}\rangle$. For real-valued defect parameters, differentiating $E_\mathrm{out} = \langle\Psi_\mathrm{out}|\Psi_\mathrm{out}\rangle$ gives

$$\begin{aligned}
\frac{\partial E_\mathrm{out}}{\partial d_n} &= \left\langle \frac{\partial\Psi_\mathrm{out}}{\partial d_n}\middle|\Psi_\mathrm{out}\right\rangle + \left\langle\Psi_\mathrm{out}\middle|\frac{\partial\Psi_\mathrm{out}}{\partial d_n}\right\rangle\\
&= 2\,\mathrm{Re}\left[\left\langle\Psi_\mathrm{out}\middle|\frac{\partial\Psi_\mathrm{out}}{\partial d_n}\right\rangle\right] = 2\,\mathrm{Re}\left[\left\langle\Psi_\mathrm{out}\middle|\frac{\partial M}{\partial d_n}\middle|\Psi_\mathrm{in}\right\rangle\right].
\end{aligned} \tag{S6}$$

Applying the chain rule to the objective function $C(\mathbf{d}) = (E_\mathrm{out}(\mathbf{d})/E_\mathrm{in} - \rho)^2/2$ in Eq. (2) of the main text, and noting that $E_\mathrm{in}$ is independent of the defect parameters, we obtain

$$\begin{aligned}
\frac{\partial C}{\partial d_n} &= \left(\frac{E_\mathrm{out}}{E_\mathrm{in}} - \rho\right)\frac{1}{E_\mathrm{in}}\frac{\partial E_\mathrm{out}}{\partial d_n}\\
&= \frac{2}{E_\mathrm{in}}\left(\frac{E_\mathrm{out}}{E_\mathrm{in}} - \rho\right)\mathrm{Re}\left[\left\langle\Psi_\mathrm{out}\middle|\frac{\partial M}{\partial d_n}\middle|\Psi_\mathrm{in}\right\rangle\right].
\end{aligned} \tag{S7}$$

Equation (3) of the main text is the multi-defect generalization of Eq. (S7).

## Supplementary Note S3. Energy suppression analysis

In the main text, we define valid solutions for the coherent-suppression target $\rho = 0$ using the criterion $E_{out}/E_{in} < 1$, because this condition corresponds to net energy suppression. We further verify that our conclusions remain valid under stricter criteria, $E_{out}/E_{in} < 10^{-1}$, $10^{-2}$, and $10^{-3}$. Figure S1a shows that the suppressed fraction decreases as the threshold becomes more stringent, but the double-defect design outperforms the single-defect design at every threshold. Figure S1b shows the grid-occupancy coverage, defined in the main text as the fraction of discretized parameter-space cells containing at least one valid solution. The double-defect coverage (solid lines) decreases with increasing $N_M$ and approaches the single-defect reference (dashed lines), confirming that the broad solution set at small $N_M$ arises from inter-defect interactions and persists across all tested thresholds.

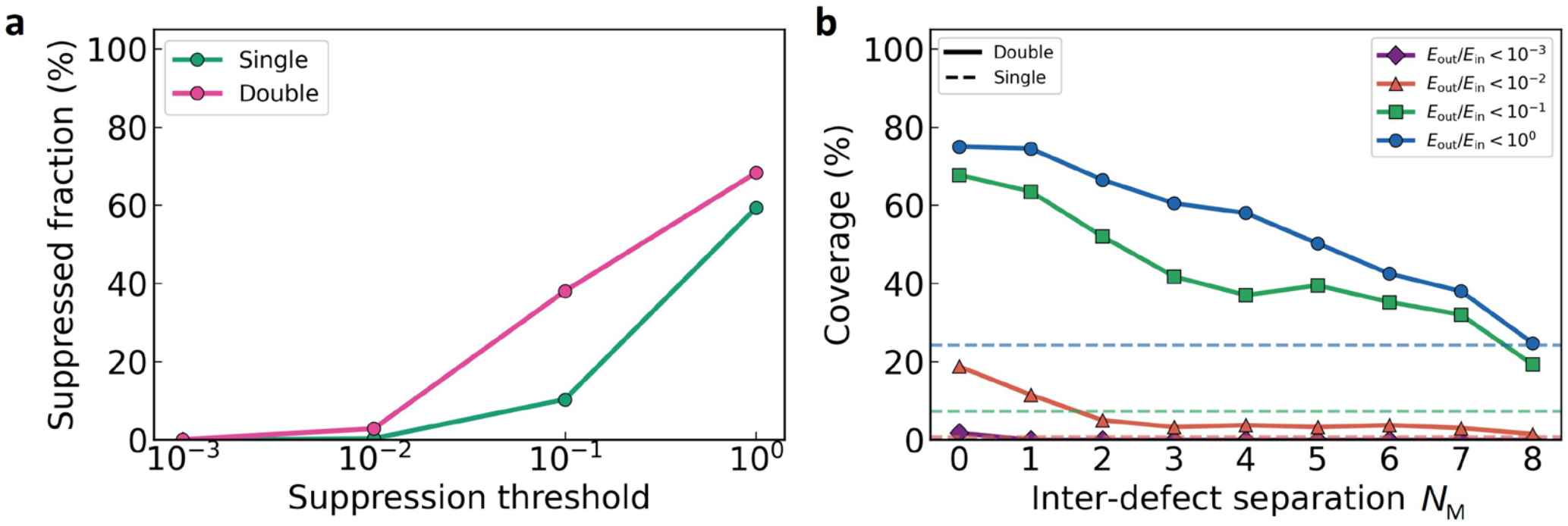

**Fig. S1. Threshold analysis. a**, Suppressed fraction as a function of the suppression threshold for single- and double-defect systems. **b**, Grid-occupancy coverage of valid solutions as a function of the inter-defect separation $N_M$ for double-defect designs. Solid lines denote double-defect coverage, and dashed lines denote the corresponding single-defect reference coverage for each threshold. The thresholds are $E_{out}/E_{in} < 10^{-3}$, $10^{-2}$, $10^{-1}$, and $10^{0}$. All other parameters are the same as those in Fig. 4 of the main text.